# EXPLORING THE OXYGEN ORDER IN Hg -1223 AND Hg -1201 BY $^{199}$Hg MAS NMR

Raivo Stern[1], Ivo Heinmaa[1], Dmitriy A. Pavlov[2], and Ingrid Bryntse[2]
[1]*National Institute of Chemical Physics and Biophysics (NICPB), 12618 Tallinn, ESTONIA,*
[2]*Department of Inorganic Chemistry, Arrhenius Laboratory, SE-10691 Stockholm, SWEDEN*

**Abstract**: We demonstrate the use of a high-resolution solid-state fast (45 kHz) magic angle spinning (MAS) NMR for mapping the oxygen distribution in Hg-based cuprate superconductors. We identify observed three peaks in $^{199}$Hg spectrum as belonging to the different chemical environments in the HgO$_\delta$ layer with no oxygen neighbors, single oxygen neighbor, and two oxygen neighbors. We discuss observed differences between Hg-1201 and Hg-1223 materials.

Key words:  $^{199}$Hg NMR; MAS; High-T$_c$ superconductors; oxygen order

## 1.  INTRODUCTION

In many of high-temperature superconductors (HTSC) the charge balance is tuned by changing the oxygen content of the system. The final result depends sensitively not just on total number of added or removed oxygen atoms but also on their distribution patterns, the details of which are impossible to study by common X-ray or microscopy methods. Nuclear magnetic resonance (NMR) spectroscopy has served as an important tool to explore HTSC allowing one to obtain valuable information about the local environment of selected atoms. We propose a high-resolution NMR technique to study the oxygen distribution in mercury containing HTSC. These materials show record high transition temperatures of 133 K[1]. Due to its ½-spin the mercury isotope $^{199}$Hg is a convenient nuclear spin label for the study of Hg based HTSC. As pointed out earlier by Hoffmann *et al.*[2], Gippius *et al.*[3], Suh *et al.*[4] and Horvatic´ *et al.*,[5] the resolution of traditional NMR measurements on randomly oriented powder samples is not high enough to record any reliable $^{199}$Hg Knight shift. Here we report first results of $^{199}$Hg NMR studies of the HgBa$_2$CuO$_{4+\delta}$ (Hg-1201) and HgBa$_2$Ca$_2$Cu$_3$O$_{8+\delta}$ (Hg-1223) phases using magic angle spinning (MAS) technique to obtain high resolution spectra.

## 2. EXPERIMENTAL DETAILS

We investigated five powder samples of Hg-1201 and Hg-1223 listed in Table 1. The Hg-1223 samples were synthesized in sealed silica tubes using an one-temperature furnace. Mixtures of oxides, with the starting nominal composition were annealed at 880 °C during 10 hours. All operations with air-sensitive reagents and products were performed in an Ar-filled glove box (MBraun Labmaster 100). The samples of Hg-1201 are as-prepared (almost optimally doped sample #4) and oxidized (overdoped sample #5).

*Table 1.* Superconducting properties of the samples: label, nominal composition, lattice constants a and c, onset of superconducting transition as measured by susceptibility (1-3) or deduced from lattice parameters[6] (4-5), isotropic shift $\sigma_{iso}$ and full width at half maximum $\Delta$ of center bands or their components as on Fig. 1.

| Nr | Nominal Composition | a [Å] | c [Å] | $T_{co}$ [K] | $\sigma_{iso}$ [ppm] | | $\Delta$ [ppm] |
|---|---|---|---|---|---|---|---|
| 1. | HgBa$_2$Ca$_2$Cu$_3$O$_{8+\delta}$ | 3.8541(1) | 15.8043(7) | 126 | A | -969(8) | 100 |
|    |    |    |    |    | B | -860(5) | 61 |
|    |    |    |    |    | C | -772(41) | 90 |
| 2. | Hg$_{0.8}$Tl$_{0.2}$Ba$_2$Ca$_2$Cu$_3$O$_{8+\delta}$ | 3.8533(1) | 15.854(1) | 129 | | -840(3) | 156 |
| 3. | Hg$_{0.8}$Sc$_{0.2}$Ba$_2$Ca$_2$Cu$_3$O$_{8+\delta}$ | 3.8593(3) | 15.806(1) | 110 | A | -982(3) | 81 |
|    |    |    |    |    | B | -868(2) | 75 |
|    |    |    |    |    | C | -756(4) | 73 |
| 4. | HgBa$_2$CuO$_{4+\delta}$ | 3.8802(1) | 9.5005(9) | 77 | | -684(9) | 237 |
| 5. | HgBa$_2$CuO$_{4+\delta}$ | 3.8751(2) | 9.4984(8) | 96 | | -653(3) | 152 |

We have made use of rather unique high-resolution solid-state NMR technique - so called fast-MAS NMR[7]. The lines of the $^{199}$Hg nuclei are typically very broad due to large chemical or Knight shift anisotropy and therefore they are hard to observe. Fast-MAS of a powder sample results in narrow resonance line at isotropic shift value $\sigma_{iso}$ (center band, marked by * on Fig. 1) and in number of spinning sidebands at multiples of spinning speed frequency. The sideband amplitudes approximately map the original broad powder pattern. In our experiments 44-46 kHz spinning speed was used. The volume of the spinning rotor (1.8 mm outer diameter) is very small, therefore, for a good quality Hg spectrum we need to average about for 24 hours. All the spectra were taken at room temperature, high spinning additionally heats the sample up by about 15 degrees. The chemical (Knight) shifts are given in Me$_2$Hg scale.

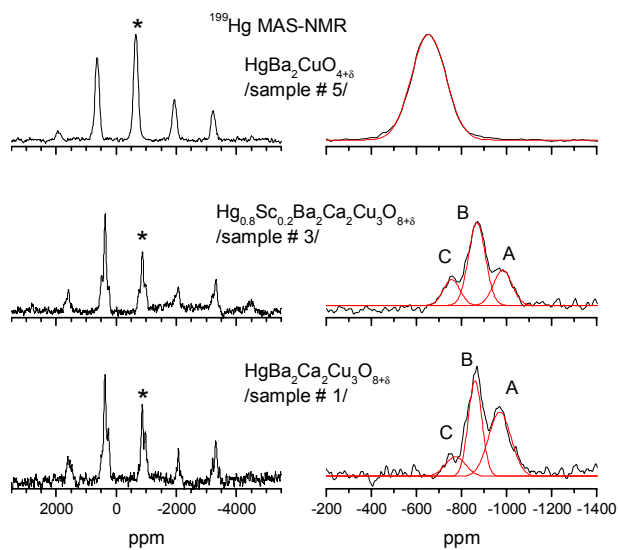

*Figure 1.* $^{199}$Hg MAS NMR spectra recorded in 4.7T field (35.79MHz). The center bands (denoted by *) of the full spectra are given in the right panels together with the Gaussian fits.

## 3. RESULTS AND DISCUSSION

The high resolution $^{199}$Hg spectra of three samples are given in Fig 1, the data for all samples are given in Table 1. In case of Hg-1201 we observe a single gaussian $^{199}$Hg line. The as-prepared sample (#4) has remarkably broader linewidth than the oxygen annealed one (#5), reflecting the higher homogeneity in the last one. In $^{199}$Hg spectra of our Hg-1223 samples (except sample #2) we observe three narrow components (labeled A, B, and C in Fig. 1). Our interpretation of the three lines is following.

According to neutron scattering results[8] there is maximally about 30% of oxygen in HgO$_\delta$ layer. Therefore the following different environments of Hg nuclei are possible
- Hg with no oxygen nearest neighbors in HgO$_\delta$ plane (assigned to line A),
- Hg with 1 oxygen nearest neighbor (line B)
- Hg with 2 oxygen neighbors (line C).

In fact the sites with 3 and 4 nearest neighboring oxygen are thinkable, but the probability to find these sites is quite low and we do not see these lines in the spectrum.

Assuming random distribution of the oxygen atoms in the layer, one can easily calculate the probabilities for all five configurations.

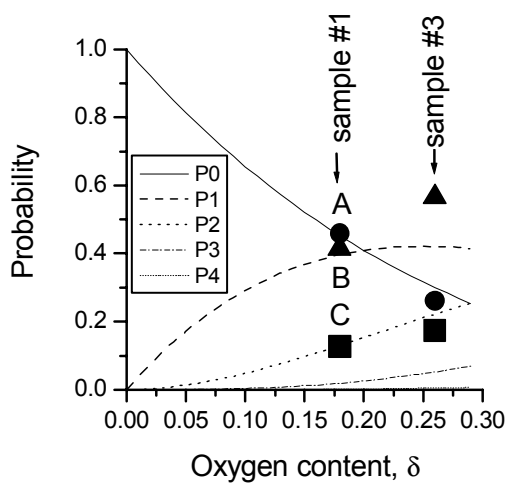

*Figure 2.* Probabilities P0, P1,..., P4 to find Hg sites with 0, 1 ...,4 oxygen neighbors in the HgO$_\delta$ layer, assuming random distribution, with the relative intensities of lines A, B, and C. The oxygen content $\delta$ for #1 and #3 is estimated from the lattice parameter a.[6]

Fig. 2 shows that relative intensities of the center band lines of the sample 1 fit well to the random scenario, whereas those of the sample 3 indicate deviation from the random distribution. Here the large intensity of the line B suggests that the oxygen atoms tend to be separated by at least one lattice period.

## ACKNOWLEDGMENTS

The financial support by the Royal Swedish Academy of Sciences, the Estonian Science Foundation, and NHMFL is gratefully acknowledged.